\documentclass[preprint,aps,pra,reprint]{revtex4-1}
\usepackage{amsmath}
\usepackage{amssymb}
\usepackage{amsfonts}
\usepackage{mathrsfs}
\usepackage{graphicx}
\usepackage{fouridx}
\usepackage{booktabs}
\usepackage{stmaryrd}
\usepackage{rotating}

\usepackage{caption}
\usepackage{subcaption}

\newcommand{\ud}{\textrm{d}}
\newcommand{\btau}{\bar{\tau}}
\newcommand{\bfk}{\mathbf{k}}
\newcommand{\bfr}{\mathbf{r}}

\newcommand{\bigbraket}[2]{\left\langle {#1} \mathrel{\left | {\vphantom {#1 #2}} \right. \kern-\nulldelimiterspace} {#2} \right\rangle}

\begin{document}

\title{Double-core Excitations in Formamide Can Be Probed by X-ray
  Double-quantum-coherence Spectroscopy }
\begin{abstract}
  The attosecond, time-resolved X-ray double-quantum-coherence (XDQC) four
wave mixing signals of formamide at the nitrogen and oxygen K-edges are
simulated using restricted excitation window time-dependent density
functional theory (REW-TDDFT) and the excited core hole (XCH)
approximation. These signals, induced by core exciton coupling, are
particularly sensitive to the level of treatment of electron correlation,
thus providing direct experimental signatures of electron core hole
many-body effects and a test of electronic structure theories.
\end{abstract}
\author{Yu Zhang}
\author{Daniel Healion}
\author{Jason D. Biggs}
\author{Shaul Mukamel}
\email{smukamel@uci.edu}
\affiliation{Dept. of Chemistry, University of California, 450 Rowland
  Hall, Irvine, California 92697, USA}
\date{\today}

\maketitle
\section{Introduction}

New attosecond X-ray table top sources\cite{popmintchev_attosecond_2010,krausz_attosecond_2009}
and the X-ray free electron lasers (XFEL)
\cite{emma_first_2010,ullrich_free-electron_2012} allow to extend nonlinear
spectroscopic techniques, originally developed to study valence excitonic
systems, to molecular core excitations.  XFEL sources
are bright enough to saturate the core-excitation transitions,
\cite{young_femtosecond_2010,hoener_ultraintense_2010} and should
enable  fundamental questions about the quasiparticle description of
many-electron states to be experimentally addressed. Nonlinear
techniques designed to probe quantum correlations in systems at lower
energies with slower dynamics may thus be extended to the X-ray
frequency regime.  The core electrons are tightly bound to specific atoms and are strongly coupled to a slower bath of
correlated valence electrons shared between atoms. 1D- and 2D-
stimulated X-ray Raman techniques have been proposed to examine
valence electron dynamics using core-holes as ultrafast switches.

This paper focuses on doubly core-excited states (DCESs), in which
the main players are the static core holes and the virtual
orbitals. These states correspond to the high-energy limit of double valence
excitations, which are challenging to treat with density functional
theory (DFT) methods.\cite{EGCM11} Multiply excited states can be
prepared by short and intense laser pulses.\cite{Connerade98} It has been
suggested that DCESs carry information about the chemical environment
of a selected
atom in double photoionization core-excitation spectroscopy.\cite{santra_X-ray_2009,ueda_extracting_2012,kryzhevoi_inner-shell_2011}
Recent experimental\cite{salen_experimental_2012,TUE12} and
theoretical\cite{kryzhevoi_inner-shell_2011,santra_X-ray_2009} work
demonstrated a static frequency shift of doubly-core-photoionized
states that depend on their local chemical environment.  The
double-quantum-coherence (DQC) signal\cite{kim_two-dimensional_2009}
preferentially targets
doubly excited states in an excitonic
system.\cite{mukamel_sum-over-states_2007} This signal strongly
depends on the coupling between excitons; infrared DQC signals,
which have been used to study vibrational couplings in the amide I
band,\cite{zhuang_dissecting_2005} sensitively depend on
anharmonicities, vanishing for harmonic vibrational systems.  In the visible,
DQC has been used to investigate the excitations of a dye molecule in
ethanol,\cite{nemeth_double-quantum_2010} and theoretical studies have
demonstrated a strong dependence on electron
correlations in model two and three-level semiconductor systems
\cite{mukamel_coherent_2007}. It
can monitor the breakdown of mean-field theories of electron
correlation in molecular systems. This technique has also been applied to
measure spin selected biexciton coupling using circularly polarized
pulses \cite{karaiskaj_two-quantum_2010,stone_two-quantum_2009}, and
interatomic couplings in a dilute potassium
vapor\cite{dai_two-dimensional_2012}. The X-ray version of this
technique (XDQC) is sensitive to correlation and exciton scattering in
DCESs, making it an attractive experimental test for many-body electron structure techniques for strongly correlated
systems. \cite{Orenstein12,Mahan,FTDK11} XDQC detects only systems in
which the singly core
excitations affect each other and the doubly core-excited wavefunction
may not be factorized into an outer product of two singly core-excited
wavefunctions. For uncoupled core excitations, two contributions to
the signal (see Sec. \ref{sec:dqcstheory}) with opposite
signs cancel out and the signal vanishes. Incomplete cancellation
therefore provides a sensitive measure of correlation between core
excitations.

Core-excited states can be calculated at various levels of theory. The
XDQC signal was first simulated for all nitrogen excitations of
isomers of aminophenol with a simple equivalent core approximation
(ECA) describes the effect of the core hole on the valence excited
states.\cite{schweigert_double-quantum-coherence_2008} In this study we employ restricted
excitation window time-dependent density functional theory
(REW-TDDFT),\cite{SFD03,BN07,DPN08a,BPT09,BA10,LFFL11,LKKG12} a
response formalism that incorporates the core hole valence
coupling through an exchange-correlation functional. This approach was
recently used to calculate the single core excitations and two
dimensional stimulated X-ray Raman signal of cysteine.\cite{ZBHG12}

Here we simulate the XDQC signals of formamide, a small organic
molecule containing carbon, nitrogen and oxygen, which is often used
as a model for the peptide backbone in proteins (see
Fig. \ref{fig:levelscheme}). Its X-ray spectra has been investigated both
experimentally\cite{ottosson_electronic_2008,ottosson_electronic_2008_2,ikeura-sekiguchi_inner_2001}
and theoretically.\cite{chong_density_2011}
Formamide has been used as a benchmark in
theoretical studies of double core hole spectroscopy\cite{AJ93,TTEY11}
and in molecular dynamics\cite{jadzyn_similarity_2012} and vibrational
spectroscopy
applications.\cite{li_two-dimensional_2008,paarmann_excitonic_2011}

We summarize the theoretical
methods and computational challenges of double excitations
in Sec. \ref{sec:quantchem}. Our treatment of core
and valence excited states is presented in Sec. \ref{sec:theory}.
Sum over states XDQC expressions for signals involving the core-excited eigenstate
frequencies and transition dipole matrix elements are given in Sec. \ref{sec:dqcstheory}. We then apply the
REW-TDDFT to calculate the orientationally averaged XANES and XDQC
signals of formamide at the N and O K-edges. The results are discussed in
Sec. \ref{sec:results}.

\section{Electronic Structure Simulations of Double Excitations}
\label{sec:quantchem}

Electronic excited states can often be represented as linear
combinations of single excitations, where a single electron is
promoted from an occupied to a virtual orbital. Some of the most
popular and inexpensive computational methods for excited states, such as
configuration interaction singles (CIS) and time-dependent density
functional theory (TDDFT), are based on this picture. While often
adequate for molecules or high bandgap materials they fail badly for
double excitations in conjugated or metallic systems with strong
valence band correlations. Double-exciton states, which are common in
molecular crystals\cite{JR66} and materials with strong spatial
localization,\cite{SROM11} also play a role in conical
intersections, long-range charge-transfer excitations and autoionizing
resonances.\cite{EGCM11} Low-lying excited states in polyenes with
significant double excitation character are notable examples in which
the single excitation picture fails qualitatively to describe the
system.\cite{SMNL93,NNH98,HHHG01,SWSD06,SWD09,KSWD10,AP11,ST12}  Strong
double-excitation features in X-ray absorption near edge structure
(XANES) spectroscopy have been reported for ferrocene and
ferrocenium compounds,\cite{OKU09} and it has been shown
that double or higher order excited configurations are necessary to
construct the spin-symmetry-adapted wave-functions of molecules with
open-shell ground states.\cite{Casida05,Casida06,TGR07,LL10}

Formally any state with energy close to the sum of two single
excitation energies can be considered a double excitation. However
this may not be always justified. The doubly excited
states in polyenes mentioned above are examples where this assignment
completely breaks down, since they are lower in energy than any single excited
state.\cite{SWSD06}
To define double excitations, we must first specify the
reference single-particle theory and define the orbitals and
their energies. In some cases a Hartree-Fock ground state reference
suggests a significant double excitation character for an excited
state while a Kohn-Sham ground state reference does
not.\cite{Casida06,EGCM11} Various high-level \emph{ab initio} methods
also show very different double excitation
character,\cite{KNNH04,SWSD06} since DCESs are directly related to
correlation. The XDQC technique creates DCESs by using two
pulses. In contrast with valence shake-up excitations in XANES induced by
the core hole Coulomb interaction, in two photon core excitation each
core is prepared by a different photon.  In this paper we
consider the valence relaxation in the field of the first core hole
explicitly, and neglect any accompanying Auger, nuclear or electronic
processes within the very short duration of the measurement.

High-level \emph{ab initio} techniques are usually required to handle
double excitations in valence band excitations.\cite{CA11} These include
complete active space self-consistent field (CASSCF), complete active
space perturbation theory of second order (CASPT2),\cite{AP11} coupled
cluster (CC),\cite{LSSJ09} multireference configuration interaction
(MRCI),\cite{MG08,ST12} symmetry-adapted cluster configuration
interaction (SAC-CI),\cite{MTM08} algebraic diagrammatic construction
(ADC)\cite{SWSD06,SWD09,KSWD10} and multireference M{\o}ller-Plesset
perturbation theory (MRMP)\cite{NNH98,KNNH04}. These accurate techniques are
computationally expensive and limited to small molecular systems such
as polyenes with several carbon atoms.

TDDFT\cite{TDDFT06,Ullrich12} balances accuracy and
computational cost for excited states. Most implementations invoke the
adiabatic approximation, by assuming that the exchange-correlation (XC)
kernel, the second functional derivative of the XC energy with respect
to density, is frequency independent. Maitra and
coworkers\cite{MZCB04} showed that a frequency-dependent XC kernel is
necessary to correctly describe a state with a strong double
excitation mixing. Double excitation energies using the adiabatic form
of the quadratic response are simply the sums of two single excitation
energies,\cite{TCM98,TC03,EGCM11} unlike early claims to the
contrary.\cite{GDP96} These trivial double excitation energies may not be found within the Tamm-Dancoff
approximation.\cite{HHG99b}  Maitra et al. also proposed dressed
TDDFT\cite{MZCB04} to remedy this deficiency; deriving a
frequency-dependent XC kernel in which single excitations are mixed
with spectrally isolated double excitations.  This approach has been
applied and tested.\cite{CZMB04,MW09,MMWA11,HIRC11} Its major
weakness is the need to assign the mixed single and double excitations
\emph{a priori}.

Many additional density functional methods have been
proposed for doubly-excited systems. Casida had added non-DFT
many-body polarization propagator correction based on the
Bethe-Salpeter equation to the XC kernel.\cite{Casida05} This
nonadiabatic XC kernel includes the dressed TDDFT kernel as a special
case if the ground state has a closed shell. Additional
frequency-dependent XC kernels accounting for double excitations in
finite and correlated systems were derived recently from the
Bethe-Salpeter equation.\cite{RSBS09,SROM11} The spin-flip (SF)
approach\cite{Krylov01a,Krylov01b,KS02,SK02,SHGK03} is also a
promising method for generating double excited states from a triplet
single reference state, in which electrons are excited to an orbital
with a different spin. The SF approach can be employed to both
wave-function-\cite{Krylov01a,Krylov01b,KS02,SK02} and
TDDFT\cite{SHGK03} based methods. In the original implementation of
SF-TDDFT, the coupling of SF excitations entered the linear response
equation only through the Hartree-Fock exchange component in the
hybrid XC functionals. Wang and Ziegler overcame this limitation by a
noncollinear formulation of the XC potential.\cite{WZ04} Recent
assessment and applications of SF-TDDFT are given in
Refs. \onlinecite{RVA10,MG11,LL12,BSK12}. The real-time approach has
also been used to obtain double excitations.\cite{IL08,LIL09} A
time-independent DFT method for multiple excitations was proposed
recently.\cite{GA10,GA12} In this approach multiple excited states
were obtained from an optimized effective potential (OEP) eigenvalue
equation with orthogonality constraints. Computing of the OEP of a
polyatomic molecule, however, is not easy, and there are still many
open questions in the DFT response formalism for excited states.
Time-dependent density matrix functional theory (TDDMFT), proposed by
Giesbertz et al.,\cite{GBaG08} may also account for double
excitations. Adopting these methods to better describe the processes
that contribute to various nonlinear spectroscopy techniques, and design experiments which test the assumptions underlying them,
is an ongoing open challenge.

\section{Theory}\label{sec:theory}

\subsection{Survey of  Approximations for Single Core Holes }\label{sec:corehole}

\begin{figure}
\includegraphics[width=5.5cm]{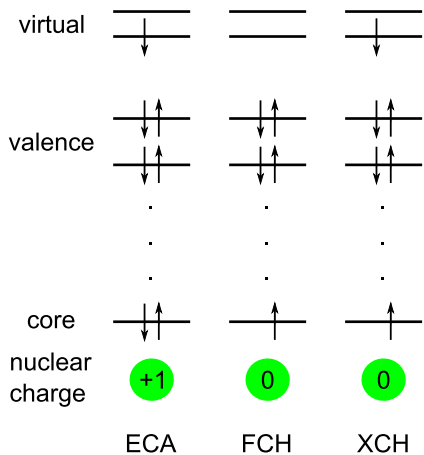}
\caption{Schematics of self-consistent calculation of
  core-excited states.  The ECA (left) replaces a core with the next
  highest element in the periodic table, the FCH (middle) fixes the
  occupation of the core-orbital, neglecting the excited electron, whereas
  the XCH (right) fixes the occupation of both orbitals. }
\label{fig:CH}
\end{figure}

The three most common approximate descriptions of core excitations
are illustrated in Fig. \ref{fig:CH}.
The equivalent core hole approximation (ECA), also known as the $Z+1$
approximation,\cite{ND69,SB76} replaces the core hole by an additional
nuclear charge, and was employed in our previous simulation studies of
X-ray nonlinear spectroscopy signals.
\cite{SM07,SM07a,SM08,SM08a,HSM08} The ECA can be easily
implemented within the routine features of standard quantum chemistry
packages, and easily represents DCESs using an extra two nuclear
charges. However, it is only applicable to deep core-holes as it neglects
the effect of the chemical environment, changes the spin symmetry of a
single core hole state \cite{WC02} and cannot account for core hole delocalization and
migration.

The full core hole (FCH) approximation
\cite{CONN05} improves on the basic ECA model by employing orbitals
determined self-consistently using the fixed core hole configuration
as shown in the middle panel of Fig. \ref{fig:CH}. The direct static
exchange (STEX) model \cite{ACVP94,ACVP97} is one type of FCH with the
occupied Hartree-Fock orbitals and the improved virtual orbitals
(IVO)\cite{HG69} of the ionic ($N-1$)-electron system. Unlike the ECA,
STEX includes orbital relaxation and can be applied to both deep and
shallow core-holes. Interactions between the excited electron and the
other $N-1$ inner electrons, however are still neglected. We have used
STEX to study linear and nonlinear X-ray spectroscopy signals
\cite{HWM11,BZHM12,ZBHG12} in small organic molecules. A detailed
analysis of STEX and comparison with REW-TDDFT is given in
Refs. \onlinecite{HWM11} and \onlinecite{ZBHG12}. Current implementations of
STEX are limited to single core hole states. Spin coupling of the two core electrons
complicates the DCES wave-function, making it hard to map to
an effective single particle Hamiltonian like that in STEX. Self-consistent
field (SCF) calculations of DCESs are also numerically tricky, and
frequently fail to converge.
Previous calculations of double core hole states using CASSCF,
\cite{ETLE10,TTEY11,TUE12}
MRCI,\cite{,KSC11,IBGG12} and ADC \cite{SKC09,KSC11,SCW83,SB84} are
expensive and had only been applied to small molecules
such as $\text{NH}_3$ and $\text{CH}_4$. Relativistic corrections may
be necessary for core electrons.\cite{KSC11,NNAA11} The much cheaper $\Delta$SCF method,
\cite{TTEY11b,TYNU11,KSC11,NNAA11} like STEX, suffers from the SCF convergence
problem for core hole states \cite{GBG08} . Moreover, running many SCF
calculations with combinations of core-holes and excited electrons is tedious.

We employ a third approach, REW-TDDFT, first proposed by Stener et al. in
2003\cite{SFD03} and further developed in additional
studies.\cite{BN07,DPN08a} REW-TDDFT only considers electrons excited
from a defined set of relevant orbitals (the restricted excitation
window), allowing to obtain high-lying states in the excitation spectrum
without calculating the lower states. A similar restricted
channel approach was suggested in Ref. \onlinecite{ENCA06}, without
orbital relaxation in the field of the core hole. Like TDDFT, the
complex polarization propagator method\cite{ENCA06} is also based on
response theory. With this method, the absorption of the system at a
given frequency can be calculated by solving a response matrix
equation; the interesting energy region can be sampled without solving
explicitly for the excited states. This method was applied in
Ref. \onlinecite{LSRA11}.

We have recently found \cite{ZBHG12} REW-TDDFT to be more accurate in
predicting frequency splitting in XANES and computationally less
demanding than STEX. Brena et al. had pointed out that TDDFT core
excitation energies have larger absolute
errors than those from STEX or transition-state calculations, because
TDDFT does not account for orbital relaxation and self-interaction of
core electrons.\cite{BSA12} We found the same trend.\cite{ZBHG12} Apart from an
overall shift, TDDFT core excitation energies agree with the XANES
of many systems very well.\cite{NFS11,LFFL11,LKKG12} The TDDFT
core-edge energy can be improved by applying the Perdew-Zunger
self-interaction correction scheme,\cite{PZ81,TRVA07} or by employing
recently-developed core-valence hybrid
functional,\cite{NION06,NIN06,IKN11} long-range corrected hybrid
functional with short-range Gaussian attenuation\cite{SWNH08} and
short-range corrected functionals.\cite{BPT09,BA10} These techniques yield core-edge energies with errors less than 1eV on a
series of small molecules.\cite{BPT09,BA10} More precise benchmarks will be
needed to compare the accuracy of STEX and REW-TDDFT.

\subsection{Double-core Excitations}
\label{sec:doubleex}

Adiabatic TDDFT cannot deal with double excitations.\cite{MZCB04}
Practical frequency-dependent XC kernels for medium or large
molecular systems have yet to be developed. Both TDDFT and the excited state core hole (XCH)
method\cite{PG06} treat singly core-excited state (SCES) very
well. The excited electron in XCH approximation is included
self-consistently through a full core hole state. XCH has been applied
to solid state X-ray absorption spectra with pseudopotentials, and
applying it to molecular system poses no additional difficulty. We
will use XCH in this paper. Pseudopotentials can be constructed for
single\cite{TCFM02} and double core hole states. An effective
potential which ignores the polarizability of the core hole in
different chemical environments will introduce errors into practical
calculations. A fixed pseudopential of a core hole, like the ECA,
ignores all core hole dynamics.

Here we employ an approach that combines XCH
and REW-TDDFT. We first run an SCF calculation to get a reference
state SCES0, with a core hole and an excited electron. This reference
is then used to run a REW-TDDFT calculation to obtain excited states
with two core-holes and two excited electrons. Adiabatic TDDFT only
treats single excitations of the SCES0 reference, not all double core
excited states are accounted for. Additional DCESs can be found by starting
REW-TDDFT at different SCES references, obtained by permuting the
occupied and virtual orbitals of SCES0 (orbital approximation as in
Ref. \onlinecite{HWM11}).

Unrestricted reference-based TDDFT is known to suffer from spin
contamination.\cite{Casida06,ICDC09} A truncated rank of
excitations from one component of a spin multiplet
generates an incomplete configuration space for the total spin operator
$\hat{S^2}$.\cite{LL10,LLZS11,LL11} Stated differently, higher
rank excited configurations are needed to represent a pure spin state. Our
current TDDFT calculations on a spin symmetry-broken reference state
are no exception, but since the singlet-triplet energy splitting is
negligible compared to the large energies of core excitations, spin
contamination does not strongly affect the calculated core excitation
energies. Casida and coworkers suggested to use the difference of the
total spin between excited states and the reference state ($\Delta
\langle S^2\rangle$) in order to filter out unphysical
excited states.\cite{ICDC09} We follow his scheme by only including
excited states with small $\Delta \langle S^2\rangle$s in our
calculations.

The two sequential core
excitations cannot be treated in the same fashion.
After the initial excitation, the valence electrons relax
self-consistently (XCH) in the field of the core hole, and the second
excitation is treated using response theory (REW-TDDFT). This
treatment uses two potentials to perturb the valence band during the
second excitation: a strong, long duration core hole potential which
must be treated nonperturbatively, and the ultrafast second photon
excitation, which can be truncated at first order using response
theory. This physically intuitive order in which relaxation precedes
photon excitation strongly effects the signal, as we show later by
reversing it, treating the original first core excitation using REW-TDDFT and
the original second using XCH.


\section{The DQC Signal}
\label{sec:dqcstheory}

\begin{figure}
  \includegraphics[width=8.5cm]{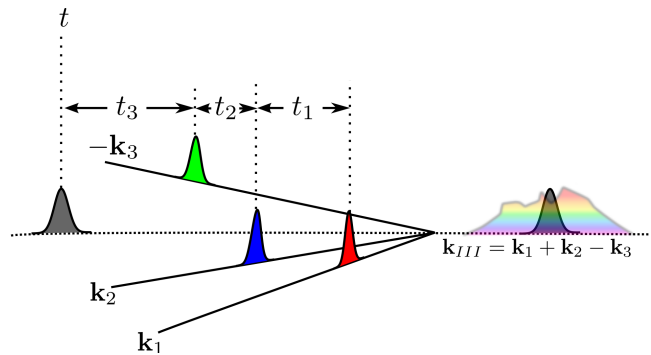}
  \caption{The XDQC technique.}
  \label{fig:xdqcpulses}
\end{figure}
\begin{figure*}
  \includegraphics[width=14cm]{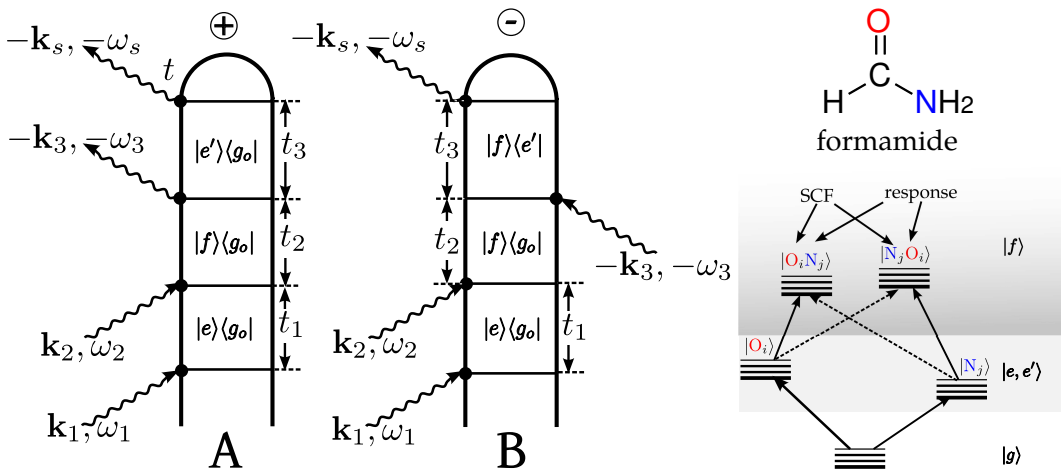}
  \caption{(left) The two diagrams contributing to the double quantum
    coherence signal. (right) Molecular structure and REW-TDDFT level scheme.  $\vert O_j N_i
    \rangle$ refers to the double core-excited states with the O1s
    electron excited to the $j$th virtual orbital of the
    self-consistent core state, while the N1s to $i$th orbital
    excitation is obtained through response theory, as described in the
    text.}
  \label{fig:kIIIloopdiags}
  \label{fig:levelscheme}
\end{figure*}

The DQC signal employs four pulses with wave vectors
satisfying $\bfk_{\text{III}} = \bfk_1 + \bfk_2 - \bfk_3$, and the transmission change of the
$\bfk_{\text{III}}$ pulse is recorded versus the three
interpulse delays ($t_1$,$t_2$,$t_3$) (as shown in
Fig. \ref{fig:xdqcpulses}).
The applied electric field is
\begin{equation}
  E(t) = \sum_{j=1,2,3,4} E_j(t-\btau_j) + \textrm{c.c.}.
\end{equation}
We represent the pulses as
\begin{equation}
  E_j(\bfr, t-\btau_j) =
  \varepsilon_j(t-\btau_j)
    \exp \left[ i \bfk_j \cdot \bfr - i \omega_j (t-\btau_j)\right]
\end{equation}
with central frequencies $\omega_j$, wave vectors $\bfk_j$ ($\bf k_4=\bf k_{\text{III}}$), and
Gaussian envelopes
\begin{equation}\label{eq:timeenv}
  \varepsilon_j(t-\btau_j) =
  \frac{
    E^o_j
    \exp \left[-(t-\btau_j)^2/2 \sigma^2_j \right]
  }{
    \sigma_j \sqrt{2 \pi}
  },
\end{equation}
with amplitudes $E^j_o$, temporal widths $\sigma_j$ and envelope
centers $\btau_j$.  The XDQC signal depends on the three (positive)
delay times ($t_j = \btau_{j+1}-\btau_j$) or
their Fourier conjugates ($\Omega_1$, $\Omega_2$, $\Omega_3$).
Invoking the rotating wave approximation and assuming
temporally well-separated pulses,\cite{schweigert_simulating_2008} the
signal is given by two the loop diagrams of
Fig. \ref{fig:kIIIloopdiags} \cite{biggs_coherent_2012,combescotprb2}
\begin{widetext}\begin{equation}
  \label{eq:siiisigtot}
  S_{\textrm{III}}(\Omega_3,\Omega_2,\Omega_1) =
  S_{\textrm{III};\text{A}}(\Omega_3,\Omega_2,\Omega_1) +
  S_{\textrm{III};\text{B}}(\Omega_3,\Omega_2,\Omega_1),
\end{equation}
where
\begin{multline}
  \label{eq:siiisiga}
  S^{\nu_4 \nu_3 \nu_2 \nu_1}_{\textrm{III};\text{A}}(\Omega_3,\Omega_2,\Omega_1) = \\
\sum_{fe'e}
  \frac{
    (\mu^{\nu_4}_{ge'} \varepsilon^*_4(\omega_4 - \omega_{e'g}) )
    (\mu^{\nu_3}_{e'f} \varepsilon^*_3(\omega_3 - \omega_{fe'}) )
    (\mu^{\nu_2}_{fe} \varepsilon_2   (\omega_2 - \omega_{fe}) )
    (\mu^{\nu_1}_{eg} \varepsilon_1   (\omega_1 - \omega_{eg}))}
  {
    (\Omega_3 - \omega_{e'g} + i \gamma_{e'g})
    (\Omega_2 - \omega_{fg} + i \gamma_{fg})
    (\Omega_1 - \omega_{eg} + i \gamma_{eg})
  }
\end{multline}
and
\begin{multline}
  \label{eq:siiisigb}
  S^{\nu_4 \nu_3 \nu_2 \nu_1}_{\textrm{III};\text{B}}(\Omega_3,\Omega_2,\Omega_1) = \\
-\sum_{fe'e}
  \frac{
    (\mu^{\nu_4}_{e'f} \varepsilon^*_4(\omega_4 - \omega_{fe'}) )
    (\mu^{\nu_3}_{g e'} \varepsilon^*_3(\omega_3 - \omega_{e'g}) )
    (\mu^{\nu_2}_{fe} \varepsilon_2   (\omega_2 - \omega_{fe}) )
    (\mu^{\nu_1}_{eg} \varepsilon_1   (\omega_1 - \omega_{eg}))}
  {
    (\Omega_3 - \omega_{f e'} + i \gamma_{fe'})
    (\Omega_2 - \omega_{fg} + i \gamma_{fg})
    (\Omega_1 - \omega_{eg} + i \gamma_{eg})
  }.
\end{multline}\end{widetext}
Here, $\varepsilon_j(\omega)$ is the spectral envelope of the $j^\mathrm{th}$ pulse
(given by the Fourier transform of Eq. \ref{eq:timeenv}), $\nu_{1 \dots 4}$ are the tensor components of the transition dipole
$\mu_{rs}$, $\omega_{rs}$ is the transition frequency
between the states $r$ and $s$, and $\gamma_{rs}$ is
phenomenological parameter describing the inverse lifetime of the
core excited state.
Different experimental techniques measure
various projections of the full three-dimensional response in
Eq. \ref{eq:siiisigtot}. Since XDQC resonances show up along
$\Omega_2$ we shall display two 2D projections of this 3D signal
\begin{equation}\label{eq:st3fixed}
  S_{\textrm{III}}(t_3,\Omega_2,\Omega_1) = \int^{\infty}_{-\infty} e^{-i \Omega_3 t_3}
  S_{\textrm{III}}(\Omega_3,\Omega_2,\Omega_1) \ud \Omega_3.
\end{equation}
where we hold $t_3$ fixed, and
\begin{equation}\label{eq:st1fixed}
  S_{\textrm{III}}(\Omega_3,\Omega_2,t_1) = \int^{\infty}_{-\infty} e^{-i \Omega_1 t_1}
  S_{\textrm{III}}(\Omega_3,\Omega_2,\Omega_1) \ud \Omega_1
\end{equation}
for a fixed $t_1$. Fixed, nonzero $t_{1,3}$ avoid the possibility of
overlapping pulses contributing to the signal. There are three
independent tensor components of the signal in isotropic media which
depend on contractions over different field polarization vectors. The
signal in an isotropic sample is a linear combination of various
contractions over the tensor components $\nu_{1\dots4}$ in
Eq. \ref{eq:siiisigtot}. The rotationally averaged signal with
all parallel pulse polarizations is given by Eq. \ref{eq:parallelsig}.

\section{Computational Details}
The optimized geometry of formamide was taken from
Ref. \onlinecite{TTEY11}. XCH calculations were performed by
converging the electron configuration with a designated core hole and
an excited electron on the lowest unoccupied molecular orbital
(LUMO). REW-TDDFT calculations were performed with the Tamm-Dancoff
approximation. The calculation of DCESs was described in
Sec. \ref{sec:doubleex}.  First a set of XCH-relaxed orbitals are
acquired, and the REW-TDDFT equations are solved in this basis.
Transition dipole matrix elements between singly and doubly excited
states are evaluated between Kohn-Sham determinants with nonorthogonal
orbitals using Eq. 10 from Ref. \onlinecite{ZBHG12}:
\begin{equation}
\langle\Psi_{\text{\tiny{A}}}|\hat{d}|\Psi_{\text{\tiny{B}}}\rangle=\sum_{m,n}^{N_{\text{config.}}}a_{m} b_{n}\sum_{i,j}(-1)^{i+j}d_{ij}^{mn}\text{Minor}({\bf S}^{mn})_{ij},
\label{eq:transD}
\end{equation}
where $\Psi_{\text{\tiny{A,B}}}$ are the SCES and DCES wavefunction respectively, $\hat{d}$ is the transition dipole operator,
$a_{m}$ and $b_{n}$ are configuration interaction (CI) coefficients for different excited
configurations ($m$ and $n$) of the SCES A and DCES B, respectively.
\begin{equation}\label{eq:dmateldef}
d_{ij}^{mn}=\sum_{p,q}c_{ip,m,\text{\tiny{A}}}^* c_{jq,n,\text{\tiny{B}}}\int\phi_{p}^{*}{\hat{d}}\phi_{q}{\text{d}}\tau
\end{equation}
is the transition dipole matrix between single excitation configurations $m$ and $n$,
$c_{ip,m,\text{\tiny{A}}}$ and $c_{jq,n,\text{\tiny{B}}}$ are MO coefficients for the configurations $m$ and $n$ of the SCES and DCES, respectively.
\begin{equation}\label{eq:smateldef}
S_{ij}^{mn}=\sum_{k,l}c_{ik,m,\text{\tiny{A}}}^* c_{jl,n,\text{\tiny{B}}}\int\phi_{i}^{*}\phi_{j}{\text{d}}\tau
\end{equation}
is the overlap matrix between the MOs of the configurations $m$ and
$n$ of state A and B, $\phi_{i,j}$ in Eqs. \ref{eq:dmateldef} and
\ref{eq:smateldef} are basis functions and $i,j,p,q,k,l$ are indices
for these basis functions. $\text{Minor}({\bf S}^{mn})_{ij}$ denotes
the $(i,j)$ minor of the matrix ${\bf S}^{mn}$.  All calculations were
carried out at the B3LYP\cite{Becke93,SDCF94}/cc-pVTZ\cite{Dunning89}
level using a modified version of the quantum chemistry package
NWChem.\cite{NWChem,LKKG12}

\section{Simulation Results}
\label{sec:results}

\subsection{XANES}
\label{sec:resultsxanes}
%
%
%
%
%
%
\begin{figure}
  \includegraphics[width=8.5cm]{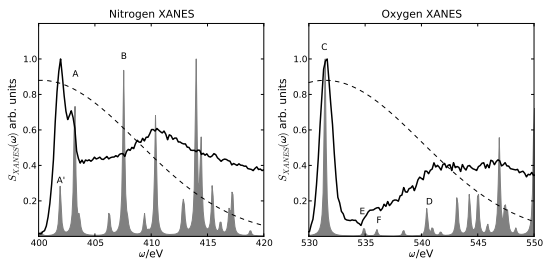}
  \caption{Calculated (grey) nitrogen (left) and oxygen (right) K-edge
    XANES for formamide. Experimental EELS
    spectra\cite{robin_fluorination_1988} are given as black lines,
    and power spectra of the pulses used in the calculation of the 2D-QCS signals as
    dashed lines. The simulated energies were shifted (+13.065 eV for
    nitrogen and +14.5 eV for oxygen K-edge) to fit the EELS
    signals\cite{robin_fluorination_1988}}
  \label{fig:rewtddftformxanes}
\end{figure}

The SCES frequencies ($\epsilon_e$) and transition dipoles
($\mu_{eg}$) were obtained using the REW-TDDFT response approach and a
ground state reference wavefunction as described in
Sec. \ref{sec:quantchem}.  $\gamma_e$, the lifetime broadening of the
core-excited state $\vert e \rangle$, is set to 0.1 eV. The XANES
signal
\begin{equation}
  S_{\textrm{XANES}}(\omega) =
    \frac{1}{\pi}\sum_{e}
  \frac{ \gamma_e \vert \mu_{eg} \vert^2}
  {(\omega - \epsilon_e)^2 + \gamma^2_e}.
\end{equation}
is shown in Fig. \ref{fig:rewtddftformxanes}.  X-ray and electron
scattering in the core energy range are known to resemble each other
in the gas phase.\cite{StohrBook} The experimental electron energy
loss spectra (EELS) are shown for comparison.

The two lowest-frequency peaks in the experimental nitrogen K-edge EELS are
split by 0.95 eV, and the low energy red peak has an intensity $\times 0.708$
relative to that of the blue peak. Our REW-TDDFT XANES
simulations show a larger, 1.29eV splitting, and a more intense blue
component ($\times 2.82$ of the red peak). The simulated oxygen edge
spectrum is much closer to experiment, and the splitting between the lowest
energy peak and higher-energy transitions is reproduced. We see a
strong, intense core-edge distinct from the higher energy shoulder.
Experimental peaks significantly higher than the ionization energy are broadened by
coupling to unbound photoelectron states, a process not included
in our simulations.

\subsection{The XDQC Signal}
\label{sec:resultsxdqc}
\begin{figure}
  \includegraphics[width=8.5cm]{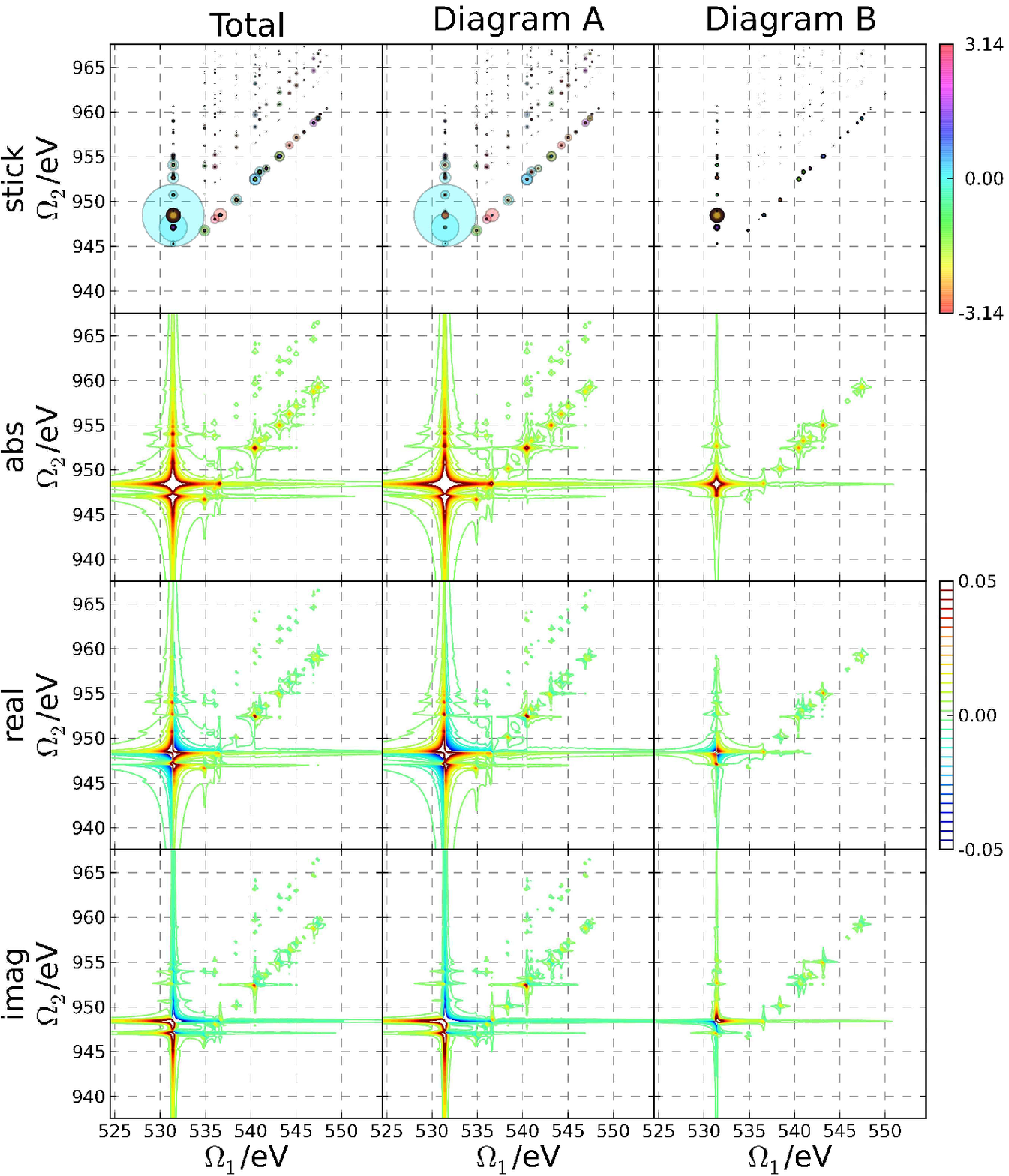}
  \caption{ The $S_{\textrm{III}}(t_3=5\text{fs},\Omega_2,\Omega_1)$
    ONNO signal with XXXX polarization configuration.  The total
    signal (left column) is the sum of the contributions from diagram
    A (middle column) and diagram B (right column) of
    Fig. \ref{fig:levelscheme}. Each circle in the stick spectra (top
    row) has a complex contribution to the signal from a combination
    of states, with the radius of the circle proportional to the
    square root of the amplitude, and colored according to the phase
    of the contributing peak. The following three rows show the
    absolute value, real and imaginary parts of the complex signal
    after convoluting with a Lorentzian of width 0.1eV. All signals
    were scaled so that
    abs($S^{\textrm{tot}}_{\textrm{III}}(t_3=5\text{fs},\Omega_2,\Omega_1)$)
    has a maximum value of one.}
  \label{fig:2dqcsomega12omega35fsONNO}
\end{figure}

\begin{figure}
  \includegraphics[width=8.5cm]{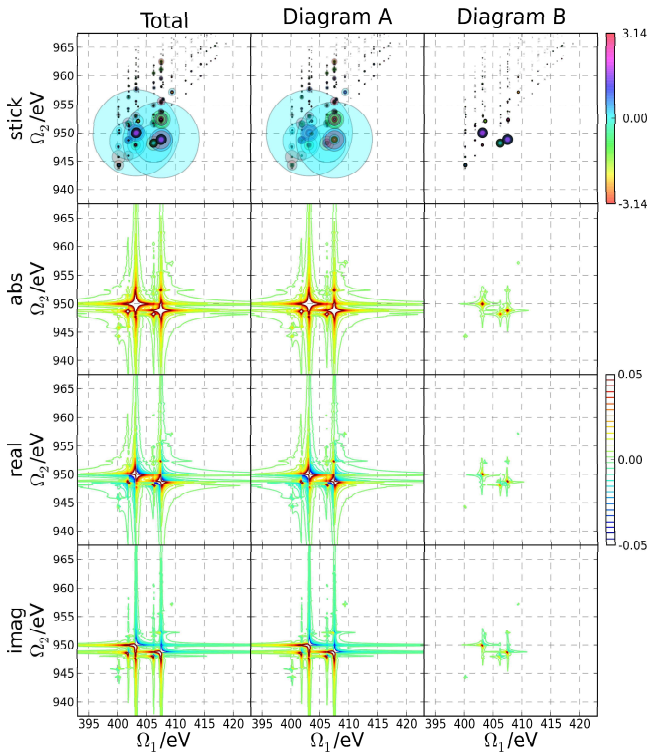}
  \caption{same as Fig. \ref{fig:2dqcsomega12omega35fsONNO}, but for the
    pulse sequence NONO.}
  \label{fig:2dqcsomega12omega35fsNONO}
\end{figure}

The
$S_{\textrm{III}}(t_3=5\text{fs},\Omega_2,\Omega_1)$ signal is
displayed in Figs. \ref{fig:2dqcsomega12omega35fsONNO} and
\ref{fig:2dqcsomega12omega35fsNONO}. In this plot, resonances on
the $\Omega_1$ axis reveal SCES ($\omega_{eg}$), and along
the $\Omega_2$ axis we see the DCES ($\omega_{fg}$) frequencies.
Time-evolution during the $t_3$ period results in a phase depending on
the final state $\vert e' \rangle$; these phases are displayed using
the colors for each peak in the 2D stick spectrum in the top row.

\begin{figure*}
   \includegraphics[width=17cm]{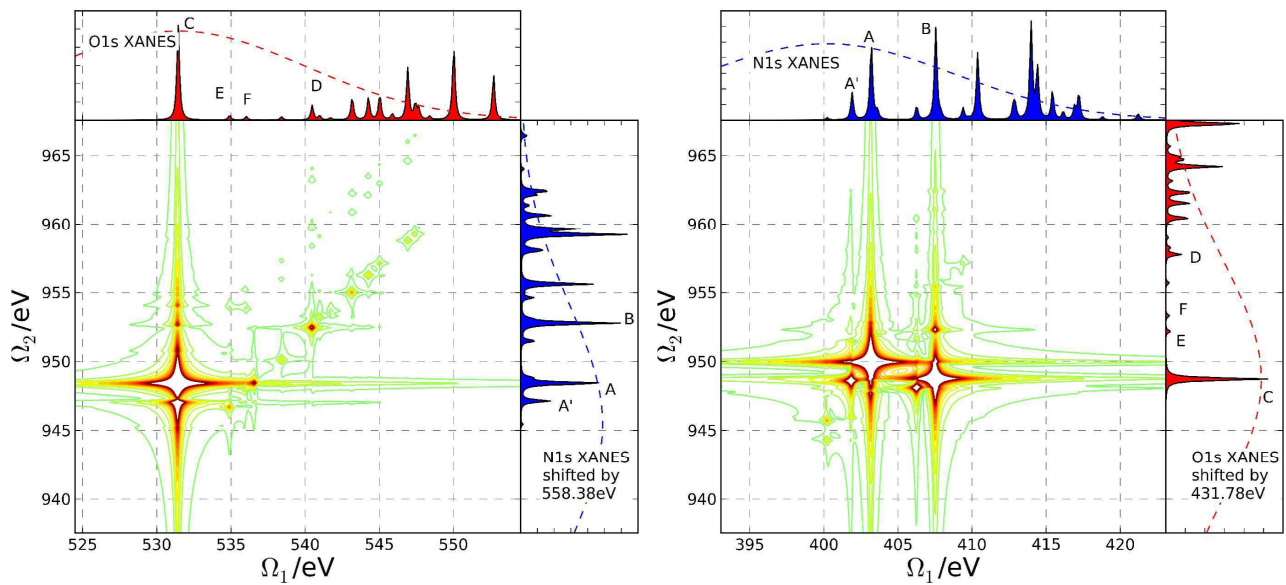}
  \caption{Comparison of the absolute parts of the ONNO (left) and
    NONO (right) $S_{\textrm{III}}(t_3=5\text{fs},\Omega_2,\Omega_1)$ signals. XANES spectra are shown in the marginals.}
  \label{fig:annotated2dxdqc}
\end{figure*}

In Fig. \ref{fig:2dqcsomega12omega35fsONNO}, we excite the oxygen core
electron first, creating resonances along $\Omega_1$, and monitor
their influence on the nitrogen core excitations, in the DCESs along
the $\Omega_2$ axis. This pulse order is marked ONNO. In the NONO
spectra shown in Fig. \ref{fig:2dqcsomega12omega35fsNONO} where nitrogen
core excitation is first, we see the influence of nitrogen core
excitations on the oxygen core excitations.  The ONNO  signal
has resonances in $\Omega_1$ and $\Omega_3$ representing
core-excitation with the same element, oxygen, in both diagram~A
($\Omega_3 \sim \omega_{e'g}$), and diagram~B ($\Omega_3 \sim
\omega_{fe'}$). Fig. \ref{fig:2dqcsomega12omega35fsONNO} shows a
typical pattern of the XDQC signal, a series of features lying on
parallel lines of roughly constant $\Omega_2-\Omega_1$ (diagonal
character). This pattern indicates weak correlation between the DCESs
and SCESs: the energies of the DCESs are roughly the sum of the
energies of two SCESs.  The oxygen core-excitations do not
substantially affect the nitrogen core-excitation spectra. In
contrast, the NONO signal (Fig.~\ref{fig:2dqcsomega12omega35fsNONO}) shows a more complicated
variation with $\Omega_1$, in which the diagonal pattern is
blurred. Two $\vert e \rangle$ resonances lie below the diagonal at
$\Omega_1 \simeq 407$ eV, suggesting that self-consistent relaxation
in the field of the nitrogen core hole after the first pulse leads to
a set of DCESs during $t_2$ in which the oxygen and nitrogen excited
states strongly interfere.  To more closely examine the difference between these
two pulse orders, we compare the peak splittings between the XANES and 2D spectra.
The energy difference between $\text{A}'$ and A is around 1.29 eV in
XANES. $\text{A}'$ and A correspond to the two strong features in
Fig. \ref{fig:2dqcsomega12omega35fsONNO} (shown more clearly in the
left panel of Fig. \ref{fig:annotated2dxdqc}), whose energy difference
is 1.34 eV, indicating that the oxygen core hole does not appreciably affect the accessible
virtual orbitals during nitrogen single core
excitation. However, the energy difference of the two strong features
on the $\Omega_2$ axis in
Fig. \ref{fig:2dqcsomega12omega35fsNONO}(shown more clearly in the
right panel of Fig. \ref{fig:annotated2dxdqc}), shows that the
dominant strong oxygen single core excitation is shifted differently
by the various nitrogen single core excitations. Comparison of the XDQC to the
XANES spectra provides useful insights on correlations between
specific core excitations.

\begin{figure*}
\centering
\includegraphics[width=17cm]{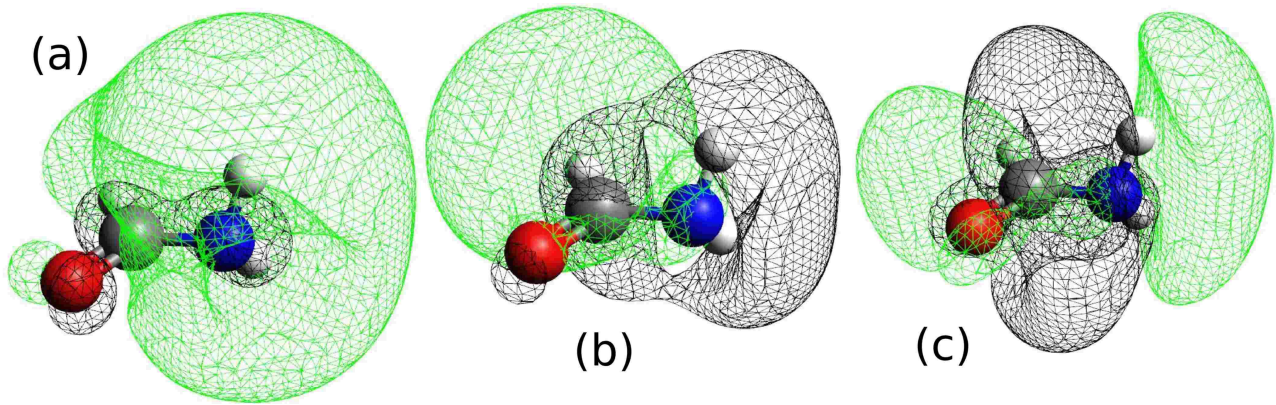}
 \caption{Dominant MOs of single particle orbitals of different SCESs discussed in Section \ref{sec:resultsxdqc}. (a)
   For peak E and $\text{A}'$. (b) For peak F (c) For peak D. Peaks are labeled in Fig. \ref{fig:rewtddftformxanes}.}
\label{fig:MO}
\end{figure*}

\begin{figure}
  \includegraphics[width=8.5cm]{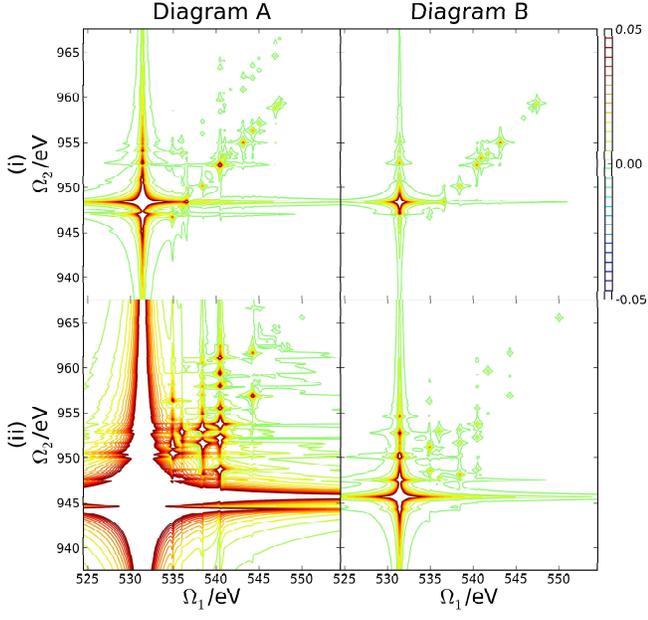}
  \caption{Comparison of the two protocols to calculate the double-core
    excited states $\vert f \rangle$ for the all-parallel ONNO
    $S_{\textrm{III}}(t_3=5\text{fs},\Omega_2,\Omega_1)$ signal as
    discussed in Sec. \ref{sec:resultsxdqc}. Contributions from
    diagram A (left column) and B (right column) with protocols i (top
    row) or protocol ii (bottom row). All graphs were multiplied by the same scaling
    factor used in Fig. \ref{fig:2dqcsomega12omega35fsONNO} }
  \label{fig:scfordercomparison}
\end{figure}

\begin{figure}
  \includegraphics[width=8.5cm]{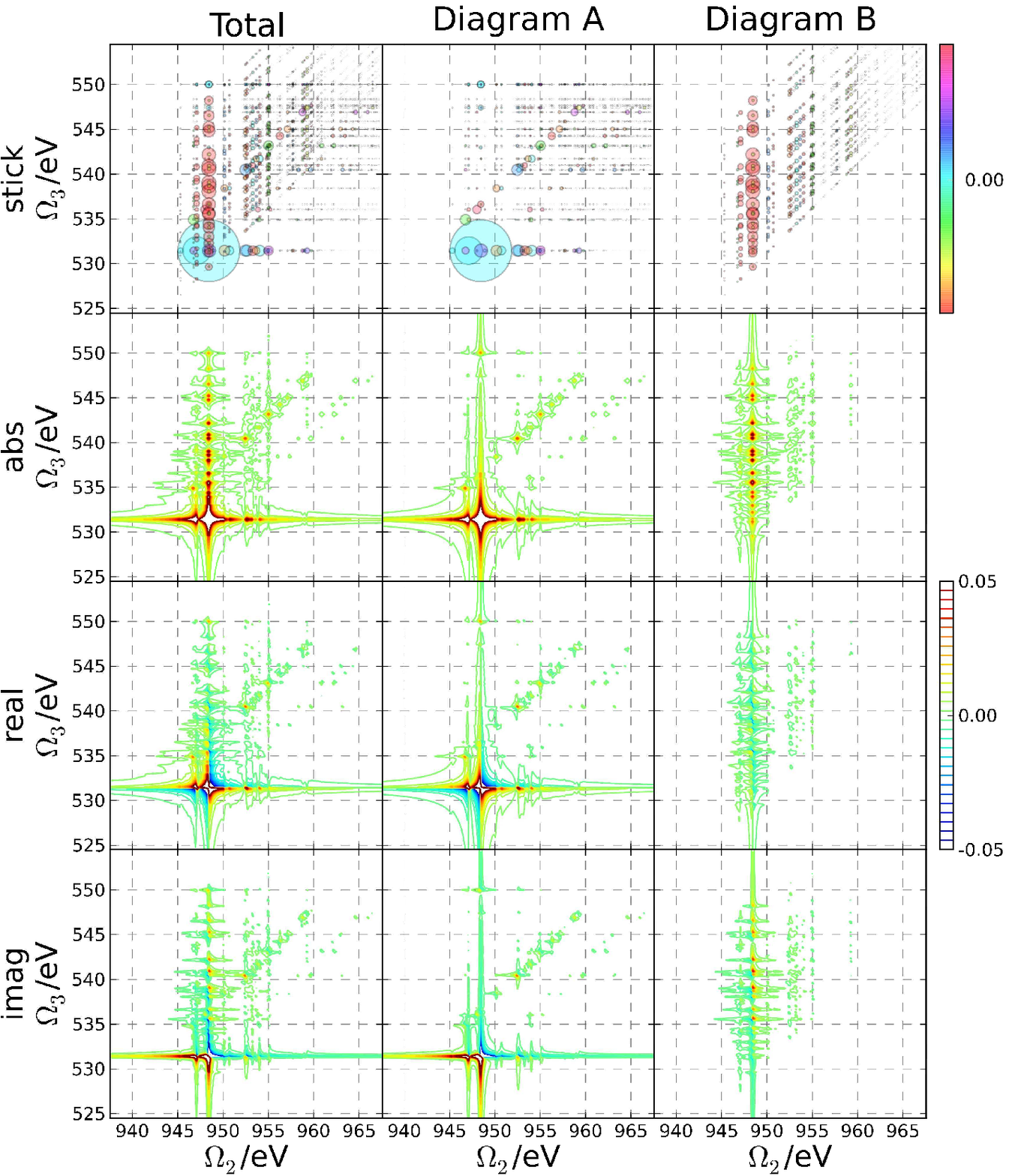}
  \caption{$S_{\textrm{III}}(\Omega_3,\Omega_2,t_1=5\text{fs})$ for the ONNO pulse
    configuration.}
  \label{fig:2dqct15fsONNO}
\end{figure}
\begin{figure}
  \includegraphics[width=8.5cm]{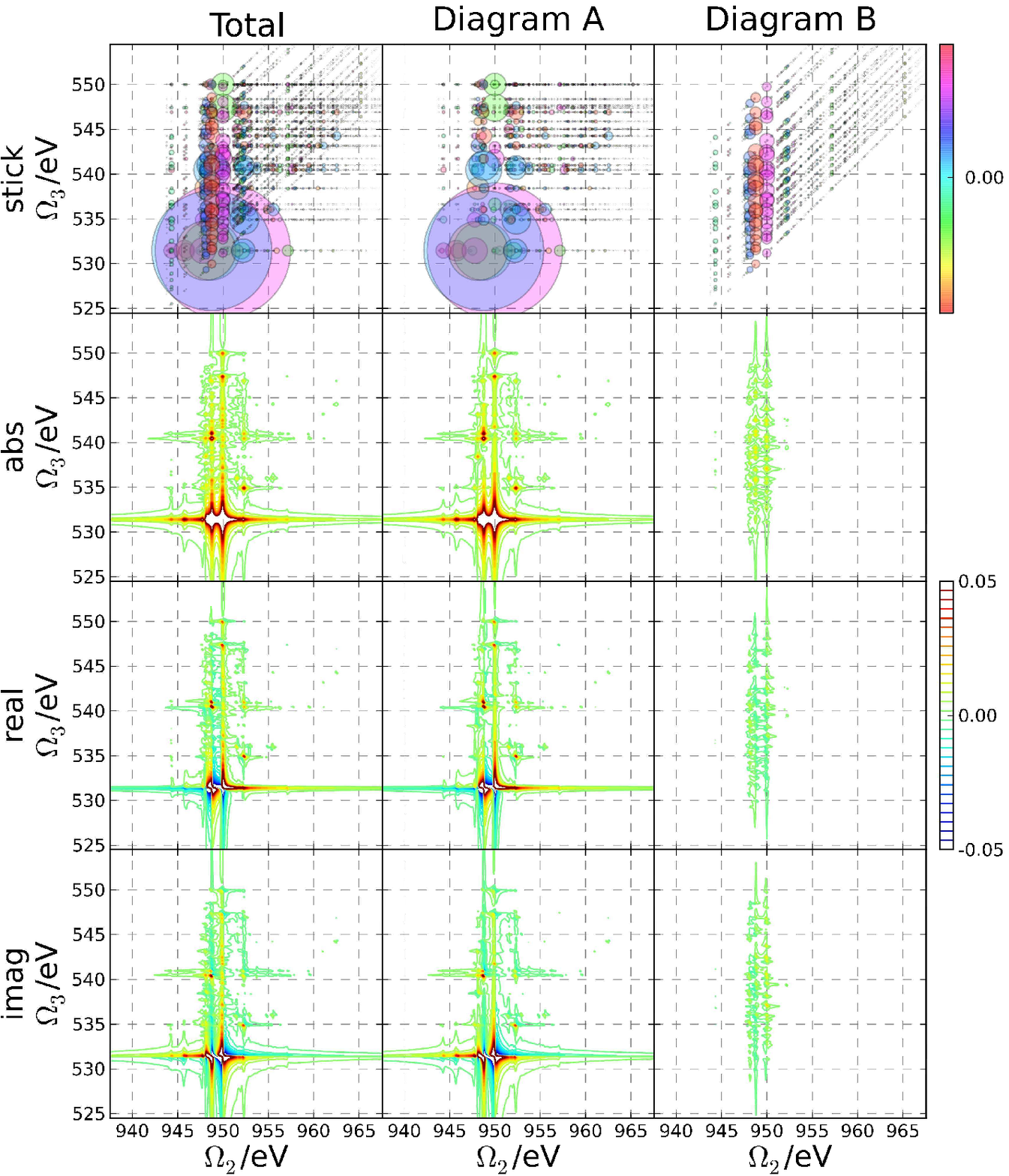}
  \caption{$S_{\textrm{III}}(\Omega_3,\Omega_2,t_1=5\text{fs})$ for the NONO pulse
    configuration.}
  \label{fig:2dqct15fsNONO}
\end{figure}

To better illustrate the relationship between the 2D-XDQC signal and
the linear absorption, we display the ONNO and NONO signals
with the XANES marginals in Fig. \ref{fig:annotated2dxdqc}.  We shift the N1s
and O1s XANES on the $\Omega_2$ axis to match the strongest
features. If the core excitations were uncoupled, the peaks in
$\Omega_2$ for a fixed $\Omega_1$ should reproduce the XANES of the
core resonant with the second pulse, with one peak quenched due to
Pauli repulsion with the previously excited core-electron. This might
explain the strong diagonal character of the ONNO signal, in which the
strong peak A is repeated for increasing $\Omega_1$. This
interpretation does not hold for the NONO signal, with its group of four
strong peaks around $\Omega_1=405$eV and $\Omega_2=950$eV. The
pattern does not match a shifted XANES signal, implying strong
correlations in the DCESs. Comparing the linear
absorption (Fig. \ref{fig:rewtddftformxanes}) to the ONNO XDQC signal
for $t_3=$5fs (Fig. \ref{fig:2dqcsomega12omega35fsONNO}), we find one
medium and two stronger features in the nitrogen K-edge XANES, and one
strong, two weak and one medium intense features for oxygen K-edge
XANES. The corresponding peaks are marked as $\text{A}'$, A, B , C, E,
F, D in Fig. \ref{fig:rewtddftformxanes}.  The N and O K-edge XANES
are well described by a single particle picture. The dominant particle
MOs in the CI expansions of these SCESs are shown on
Fig. \ref{fig:MO}.

In the ideal zero-DCES-SCES correlation case we expect to see two
parallel lines of strong features in
Fig. \ref{fig:2dqcsomega12omega35fsONNO} corresponding to A and B in
the nitrogen XANES, but we only see one clear set of ``diagonal''
resonances.  The peaks corresponding to states E and F in the oxygen
XANES are much weaker than that of D, but their corresponding peaks in
the XDQC spectra (in Fig. \ref{fig:2dqcsomega12omega35fsONNO}) are as
strong as as that of D. State E shares the same particle MO as that of
$\text{A}'$, implying a strong coupling between E and $\text{A}'$,
which explains the corresponding strong XDQC signal. But for the XDQC
peak corresponding to F, it is very difficult to compare the coupling
strength of F and $\text{A}'$ to that of D and $\text{A}'$, simply by
their dominant particle MOs. Even the calculated energy of the DCES is
close to the sum of two single excitations, the wavefunction is not
the simple outer product of two SCES wavefunctions. The single
particle picture can not explain why the XDQC peak corresponding to F
also becomes strong, suggesting that many-body effects dominate the
signal.

As discussed in Sec. \ref{sec:quantchem}, the DCES ($\vert f \rangle$)
strongly depend on the order of core excitation by the first two
pulses, showing that XDQC is sensitive to the order of the formation of doubly core-excited states. Our XDQC
simulations also show large differences between the two protocols explained in the following,
reflecting the approximate treatment of many-body effects.  In Figs. \ref{fig:2dqcsomega12omega35fsONNO} and
\ref{fig:2dqcsomega12omega35fsNONO} we assumed the $\vert f \rangle$
states were relaxed in the field of the first (core hole a), and
calculated using response theory with respect to the second excited
core (core hole b, protocol i). In Fig. \ref{fig:scfordercomparison}
we present an example of the opposite protocol for the ONNO technique:
the $\vert f \rangle$ states are generated using SCF relaxation for
core hole b and then response theory for core hole a (protocol ii).
We observe a different blurred line pattern in the spectra obtained
with protocol ii (see Fig. \ref{fig:scfordercomparison}). Different levels of theory show a different type
of DCES-SCES correlation.  The protocol i diagram A signal also shows
a stronger diagonal component than the corresponding protocol ii
signal, which displays a different density of two-particle states on
the $\Omega_2$ axis. Comparing XDQC signals from
different pulse orders with experiments should reveal how well those
theories treat the very specific correlation between a DCES and
SCES. This could be valuable in future methodology development for DCESs.

Another portion of the signal,
$S_{\textrm{III}}(\Omega_3,\Omega_2,t_1=5\text{fs})$, is shown in
Figs. \ref{fig:2dqct15fsONNO} and \ref{fig:2dqct15fsNONO}.  In these
graphs, resonances along $\Omega_2$ correspond to doubly excited
states ($\omega_{fg}$), and $\Omega_3$ shows the final single
excited states ($\omega_{e'g}$ and $\omega_{fe'}$) in the XDQC
process. As with the
$S_{\textrm{III}}(t_3=\textrm{5fs},\Omega_2,\Omega_1)$ signals, there
are diagonal characteristics in the ONNO spectra
(Fig. \ref{fig:2dqct15fsONNO}) which are absent in the NONO spectra
(Fig. \ref{fig:2dqct15fsNONO}).  This is further evidence for the
increased correlations for the states probed by the ONNO
process. In the $S_{\textrm{III}}(t_3=\textrm{5fs},\Omega_2,\Omega_1)$
signal the resonances for both A and B diagrams are the same, only the
phases for two contributions differ. The
$S_{\textrm{III}}(\Omega_3,\Omega_2,t_1=5\text{fs})$ signal has
qualitatively different peaks along $\Omega_3$ for the two diagrams, as
expected, since these diagrams differ in this time
period.

\section{Conclusions}

Typically the overlaps of valence holes or electron orbitals are higher than
those of core holes or electrons. The correlations between valence
excitations are stronger than those between core excitations, which makes
the corresponding DQC signals more complicated
than those in the present work.\cite{LAM08}
We have demonstrated qualitative differences in the stimulated
XDQC signals depending on the order of core-excitations at different
elements in a small organic molecule.  As an experimental technique,
XDQC may be used to fingerprint different theoretical approaches for
modeling core-excitation, by dissecting dynamical doubly-core-excited
resonances corresponding to the effects of valence relaxation induced
by the core hole. Additional work will be required to further
analyze the spectra reported here.  First, a detailed description of
the valence dynamics, and a comparison of the TDDFT theory with the
more easily interpreted orbital theory should highlight the effect of
electron correlation on the XDQC signal.  Second, with a higher level
of theory to describe DCESs, the variation of the $\Omega_2$
resonances with $t_1$ can be used to measure how the order of
core-excitation affects the two-particle density of states.  The
core hole has a unit charge which creates a very strong field over
atomic length scales.  If the delay between the first two pulses is
varied, the changes in the  $\vert f \rangle$ resonances along $\Omega_2$ will
reflect the order of perturbation theory necessary to correctly model
this strong interaction with the valence band.  For short $t_1$,
linear response theory alone, or supplemented with a full or partial
core hole orbital transformation may adequately capture this effect,
as it does in linear absorption. We demonstrated significant differences
between the XDQC signals predicted by the calculation protocols.
Interpolation between the extremes of linear response (first order),
and a full SCF calculation will require a higher level of theory to
describe many-body effects in the core-excitation.

In the simulations presented here the relaxation of single and double core-excited
states, is treated phenomenologically.  We
only include population decay through Auger or a radiative process
that destroys the core hole. We further use the same decay rate for
all SCESs and DCESs, independent of the orbitals involved.  Interaction
of the system with a bath, composed of vibrational and
valence-electron degrees of freedom, could introduce additional pure
dephasing. Nonlinear spectroscopy has been successful in lower
frequency regimes (NMR to the visible) at disentangling different
underlying mechanisms for excited state decay and
dephasing.\cite{mukamel_principles_1999} Similar effects are expected in
core-excitation spectra. Estimating their
magnitude, comparing them to population decay rates, and
designing experiments that distinguish between them is an interesting
future topic. One manifestation of these effects in
existing X-ray experiments are fluorescence vs. Raman signals, which
are controlled by the ratio of pure dephasing and population
relaxation rates.

\acknowledgments{ We gratefully acknowledge the support of the
  Chemical Sciences, Geosciences and Biosciences Division, Office of
  Basic Energy Sciences, Office of Science, U.S. Department of Energy,
  the support of the National Science Foundation (NSF) through Grant
  No. CHE-1058791 and the National Institute of Health (Grant
  GM-59230). Help on the REW-TDDFT calculations from Niranjan Govind of
  the Pacific Northwest National Lab (PNNL) is greatly appreciated.
}

\appendix



\section{Orientational Tensor Averaging}
\label{sec:rotavg}

The dipole and electric field polarization tensor contributions to the
signal must be rotationally averaged over all possible orientations of
the molecule.  The rotationally
averaged\cite{andrews_three-dimensional_1977} signal is
\begin{equation}
  S^{\textrm{rot}}_{\textrm{III}} = \frac{1}{30} \left(
  \begin{array}{c}
    S^{\alpha}_{\textrm{III}}\\
    S^{\beta}_{\textrm{III}}\\
    S^{\gamma}_{\textrm{III}}
  \end{array}
  \right)
  \left(
  \begin{array}{rrr}
    4 & -1 & -1 \\
    -1 & 4 & -1 \\
    -1 & -1 & 4
  \end{array}
  \right)
  \left(
  \begin{array}{c}
    \cos \theta_{12} \cos \theta_{34} \\
    \cos \theta_{13} \cos \theta_{24} \\
    \cos \theta_{14} \cos \theta_{23} \\
  \end{array}
  \right)
\end{equation}
where $\cos \theta_{ij} = \hat{e}_i \cdot \hat{e}_j$, $\hat{e}_{i}$ is the
the polarization of the $i$th pulse polarization vector, and
$S_{\alpha,\beta,\gamma}$ represent various contractions over the
tensor components of the response;
\begin{equation}
  S^{\alpha}_{\textrm{III}} = \sum_{\nu_1 \nu_2} S^{\nu_2 \nu_2 \nu_1 \nu_1}_{\textrm{III}},
\end{equation}
\begin{equation}
  S^{\beta}_{\textrm{III}} = \sum_{\nu_1 \nu_2} S^{\nu_2 \nu_1 \nu_2 \nu_1}_{\textrm{III}},
\end{equation}
and
\begin{equation}
  S^{\gamma}_{\textrm{III}} = \sum_{\nu_1 \nu_2} S^{\nu_1 \nu_2 \nu_2 \nu_1}_{\textrm{III}}.
\end{equation}
For the all parallel pulse configuration, the signal is proportional
to the sum of all three
\begin{equation}\label{eq:parallelsig}
  S^{\textrm{rot}}_{\textrm{III}} = \frac{1}{15} \left(
    S^{\alpha}_{\textrm{III}} +
    S^{\beta}_{\textrm{III}} +
    S^{\gamma}_{\textrm{III}}  \right)
\end{equation}

%

\end{document}